# Fuzzy Logic based Systems for Autonomous Parking – Part I: An Integrated Multi-functional System

Yu Wang and Xiaoxi Zhu

*Abstract—* **This paper presents an intelligent autonomous parking system with multi-functions. The integrated system consists of two sub-systems, namely the Fuzzy-Based Onboard System and control center. Unlike most current auto-parking experiments, this FBOS enables the car to perform both slot detection and parking under two different parking modes. Facilitated by the control center, real-time monitoring is also achieved to reduce chances of error.**

## I. INTRODUCTION

The past decade has seen a sharp increase in the number of vehicles. One of the side effects is the growing difficulty of parking. Searching for parking space is time-consuming. Besides, scratches and bumps often occur in parking process. A reliable autonomous-parking module is desired to resolve the trouble. Key issues in successful autonomous parking include posture stabilization, steering angle control and path planning. Various control theories have been proposed to address these issues. Continuous time-varying feedback and invariant manifold techniques are proved effective in achieving posture stabilization in nonholonomic systems([1]-[2]). The following works ([3]-[4]) present other methods to stabilize steering angle for unmanned vehicles. Path planning strategies have also been modelled in [5]-[6]. Recently, fuzzy logic control in auto-parking has received increasing attention. Controllers based on fuzzy logic, mimicking human reasoning, are inherently superiors than other controllers in handling uncertainties in the parking process([7]-[11]).

Most of the past researches on autonomous parking have a narrow focus. The auto-parking vehicles simply move into the empty space once it is placed properly in front of a parking slot. It is still dependent on human observation to find a suitable slot for parking, which cannot be completely trustworthy. In this paper, an intelligent autonomous parking system which achieves both slot detection and parking is proposed. The system is based on fuzzy logic control. Three fuzzy logic controllers are designed and implemented to tackle the issues of posture stabilization, steering angle control and parking decision-making. Parameter tuning of control rules through experiments yields stable performance.

The first author is with Yale University, yu.wang@aya.ayle.edu, the second author is working at Google now.

Large car parks are common in rural areas. Parking in such places is easier and safer due to less environmental versatility. However, it is time-consuming to find an empty slot. The intelligent autonomous parking system is designed to facilitate parking in this setting. Car owners can simply leave their car at the entrance and fetch it at the exit. The intermediate process, like searching for an empty slot, parking, and moving out are handled by the car, with aid of the control center.

Such an intelligent system has great potential in real-life application. Large companies can implement this system to reduce wasted time on parking during peak hours. Shopping centers can built such a car park to avoid congestions. Neighbourhood can also integrate this system as part of their public infrastructure.

## II. SYSTEM FUNCTIONALITY

The intelligent autonomous parking system is an integrated multi-functional system with great potential for commer-cialization. It consists of two sub-systems, the Fuzzy-Based Onboard System (FBOS) and control center. The control center contains two parts, the ground station which is located within the car park and the Graphic User Interface (GUI) on PC in the control room. Ground station communicates with the car directly through real-time duplex RF communication. GUI is connected to the ground station through RS232 serial port cable. The main purpose of GUI is to monitor the car park condition and parking process. It also supports the additional feature to communicate with car owners through text messages.

The system is able to accomplish autonomous parking without human intervention. Once a car is driven to the entrance, the owner can switch the car to auto-parking mode. First of all, a parking request is sent to the control center. Upon receiving the request signal, officer in the control center will check the car park and grant permission if no abnormalities detected. In the process of auto-parking, the car keeps sending back its current status, (i.e. searching for parking slot, parking, etc). This information is shown on the Graphic User Interface. The car slowly drives along the wall and searches for available parking slot. Once a suitable space is detected, the control center will be informed about the location and the corresponding parking mode (i.e. parallel parking or vertical parking). An end signal is sent to Control

Center once the car is properly parked. Car owner receives notification about car location via text message then. The car switches to sleep mode, with all functions disabled, except for RF communication. If the driver does not want to walk through the huge car park to fetch his car, he can send a leave request back to the control center via SMS. Upon receiving customer request, control center will forward it to the car via ground station. The car switches to leaving mode, and drives itself to the exit.

Since safety is the primary concern, the system is designed such that it is able to handle unexpected interruptions. Obstacle avoidance is controlled by FBOS. Real-time RF communication enables the car to report any abnormal situations in the car park, such as congestion, crash, and fires. Workers monitoring the car park in the control room will respond accordingly.

## III. Fuzzy Based Onboard System

### A. Overview

The key component of this autonomous parking system is the Fuzzy Based Onboard System (FBOS). The FBOS is an integrated system which enables the car to perform different functions like unmanned driving, parking slot detection, parallel/vertical parking and real-time communication. FBOS consists of the following components: microprocessor, fuzzy logic controller, sensors and RF communication module.

The car equipped with the FBOS has different action modes defined as follows:
1) Searching: actively search for a suitable parking slot
2) Parking: complete parallel/vertical parking
3) Sleep: idle state, waiting for commands
4) Leaving: automatically drive to the exit

A duplex real-time communication channel between car and the control center is maintained. The current action mode and position of the car is updated continuously. In case of emergencies, car sends alarms to the control center and wait for instructions.

### B. Sensor Arrangement

Since movements of the car are confined in one flat plane, infra-red sensors which measures linear distance are sufficient to facilitate autonomous parking. The infra-red proximity sensor consists of a transmitter and a receiver. The transmitter emits infra-red light, which bounces back when it encounters obstacles. The receiver converts the magnitude of the reflected light to voltage levels. Therefore, the voltage output is proportional to the reciprocal of detected distance. An accurate expression can be obtained from calibration.

Six sensors are used to facilitate various tasks such as posture stabilization, environment monitoring, path detection etc. The six sensors are placed at front center, back center, front left, back left, front right and back right (labelled as FC, BC, FL, BL, FR and BR). The two sensors on the right side are used to maintain posture stabilization while marching down the car park. Sensors on the left side detect size of parking slots. The front and back sensors help avoid obstacles.

### C. Fuzzy Logic Controller

First introduced by Zadeh in 1965, fuzzy logic has aroused great interest due to its similarity to human reasoning. Unlike Boolean logic which has only two states (TRUE and FALSE), fuzzy sets can have several truth values between extreme cases (0 and 1). Controllers based on fuzzy logic outperforms conventional controllers due to its superiority in handling uncertainties and imprecision.

A typical fuzzy controller consists of three stages, input stage, decision stage and output stage. In the input stage, also known as fuzzification, single-value input is mapped to a linguistic variable. A linguistic variable is a fuzzy set with pre-defined membership function. Conventional control rules define deterministic mathematical relationships between inputs and outputs. Decision making in fuzzy controller (i.e. inference), however, is based on IF-THEN rules. The output stage, usually called defuzzification, combines possible outcomes from each rule to generate a specific control value. In the proposed FBOS, fuzzy logic controller is used to achieve posture stabilization and steering, as well as decision making in slot detection.

*1) Posture Stabilization:* Posture stabilization is crucial in autonomous parking. It helps avoid collisions and ensures accurate measurement of the parking slot size. Car exhibits a wall-following behavior by steering front wheels. This is controlled by fuzzy logic controller, based on real-time measurements from sensors FR and BR (Fig.1).

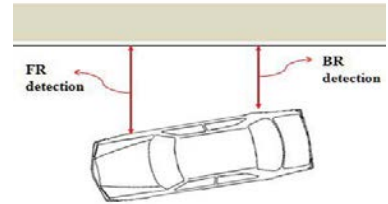

Fig. 1. Posture Stabilization

The fuzzy logic controller takes in two inputs:

$$Xd = \begin{cases} FR - LIMIT & \text{if moving forward} \\ BR - LIMIT & \text{if moving backward} \end{cases}$$

$$Xe = \begin{cases} FR - BR & \text{if moving forward} \\ BR - FR & \text{if moving backward} \end{cases}$$

Both of these inputs are fuzzified based on the following membership function. The universal set is divided into five partitions: positive large (PL), positive small (PS), zero (ZO), negative small (NS), and negative large (NL). The output of fuzzy logic controller is the steering angle of front wheels, $\varphi$, which is a crisp set. Steering front wheels to the right is defined as steering in the positive direction. The output is also divided into five sets, namely positive large (PL), positive small (PS), zero (ZO), negative small (NS) and negative large (NL). The membership functions of input and output are given in Fig.2.

TABLE I
IF-THEN RULES FOR POSTURE STABILISATION

| Xe \ Xd | NL | NS | ZO | PS | PL |
|---|---|---|---|---|---|
| NL | NL | NL | NS | NS | ZO |
| NS | NL | NS | NS | ZO | PS |
| ZO | NS | NS | ZO | PS | PS |
| PS | NS | ZO | PS | PS | PL |
| PL | ZO | PS | PS | PL | PL |

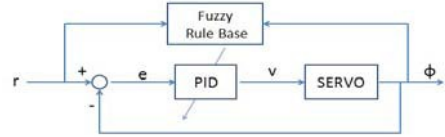

Fig. 3. Control Diagram

Before turning starts, the current angle is cleared to zero. The set point r is the desired turning of 90/-90 degrees. The output is the current angle $\varphi$. The difference between set point and current angle is the error angle e. (Fig.4)

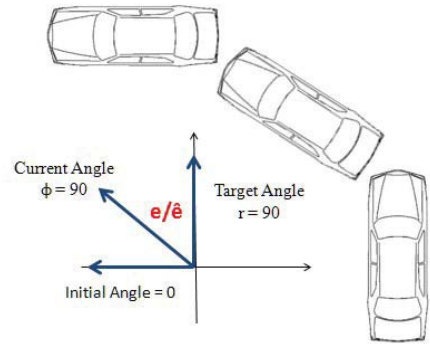

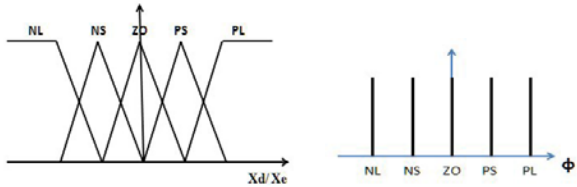

Fig. 2. Membership Function for input,output in Posture Stabilization

If both Xd and Xe are positive large, the car is far away from the wall and the distance is still increasing. Therefore, it is desirable to steer the front wheel to the right (positive direction) to make a correction. The fuzzy IF-THEN rules can be stated in the linguistic form as:

IF Xd is PL and Xe is PL, THEN $\varphi$ is PL
IF Xd is PL and Xe is PS, THEN $\varphi$ is PL
IF Xd is PL and Xe is ZO, THEN $\varphi$ is PS
......
IF Xd is NL and Xe is NL, THEN $\varphi$ is NL

The complete IF-THEN rules are summarized in Table I. Posture stabilization can be maintained via the designed control rules. Even if the initial car position is not parallel with the wall, it can automatically adjust itself.

2) *Steering control:* One critical issue in motion control is to steer the car by a fixed angle. This feature is especially important in vertical parking, where car should turn by 90 degrees. To achieve this, a fuzzy-based PID controller is implemented. The control diagram is illustrated in Fig.3.

To record the turning angle $\varphi$, ArduIMU is integrated in the FBOS. ArduIMU is an Inertial Measure Unit combined with Arduino-compatible processor. It runs the Attitude Heading Reference System (AHRS) code to calculate angle value with respect to a reference point based on angular velocity measurement from gyroscope.

Fig. 4. Angle Control

Inputs to the PID controller is the error angle and the change of error $\hat{e}$, i.e. angular velocity from ArduIMU. The output is a voltage signal to control the steering angle of DC servo. Both inputs and output are mapped to the following fuzzy sets: positive large (PL), positive small (PS), zero (ZO), negative small (NS) and negative large (NL). Membership functions are given in Fig.5. The fuzzy IF-THEN rules are summarized in Table II.

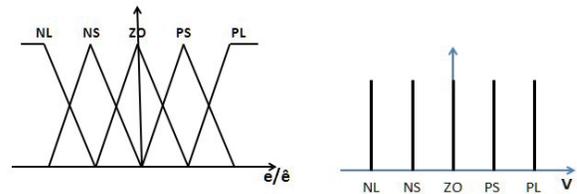

Fig. 5. Membership Function for input,output in Angle Control

3) *Parking slot detection:* The car park is designed such that all parking slots are on the left side of the parking lane. If a particular slot is occupied, the measurement of sensor FL must be a small value. If a slot is empty, sensor FL reports a large measurement. Therefore, whether a slot is empty can be easily determined.

Both width and length of the parking slot should be considered while making a decision on parking mode. The

TABLE II
IF-THEN RULES FOR STEERING CONTROL

| ė\e | NL | NS | ZO | PS | PL |
|---|---|---|---|---|---|
| PL | ZO | PS | PL | PL | PL |
| PS | NS | ZO | PS | PL | PL |
| ZO | NL | NS | ZO | PS | PL |
| NS | NL | NL | NS | ZO | PS |
| NL | NL | NL | NL | NS | ZO |

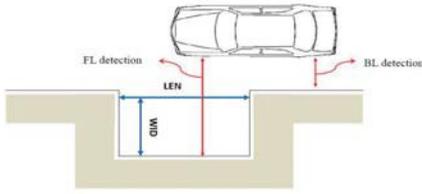

Fig. 6. Slot Detection

width of parking slot (WID) corresponding to the measurement of sensor FL, adjusted by a small offset (illustrated in Fig.6). The length, however, is calculated using the following method. A rotary encoder is fixed on the axle connecting two front wheels to count number of wheel rotations. Whenever the measurement of sensor FL switches from small to large, the rotary encoder is cleared to zero. As soon as the measurement switches from large to small, value in rotary encoder is read out. Therefore, the length of the empty slot is given by $LEN = 2 \times \pi \times diameter \times counter$

Two inputs to the fuzzy logic controller are the slot dimensions WID and LEN that can be mapped to any one of the three fuzzy sets: suitable for parallel parking (PS), suitable for vertical parking (VS), and not suitable for parking (NA). Outputs are discrete sets indicating the next step actionn, namely parallel parking (P), vertical parking (V), and continue searching (C). The membership functions are shown in Fig.7.

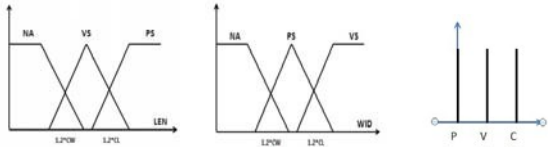

Fig. 7. Membership Function for inputs,output in Parking Decision

The IF-THEN rules are summarized in Table III.

### D. Parking

Based on the length and width of the empty slot, two parking modes are possible: parallel parking and vertical parking. The steps for parallel parking are as follows (Fig.8):
1) Adjust initial position

TABLE III
IF-THEN RULES FOR PARKING DECISION

| WID\LEN | NA | VS | PS |
|---|---|---|---|
| NA | C | C | C |
| PS | C | C | P |
| VS | C | V | V |

2) Steer to the left and move backward
3) Steer right and move backward
4) Post-parking adjustment inside the slot

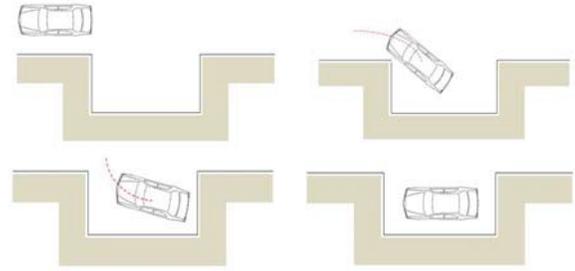

Fig. 8. Parallel Parking

The steps for vertical parking are as follows (Fig.9):
1) Adjust initial position
2) Backtracking turn 90 degrees
3) Move backward straightly

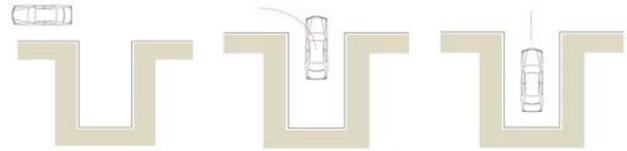

Fig. 9. Vertical Parking

### E. Emergency Handling

Obstacle-avoidance is handled by FBOS. Car stops immediately whenever an obstacle is detected and resumes moving if it is cleared. A timer starts counting when car stops. If the car does not resume moving after a certain time limit, a signal indicating congestion will be sent to the control center. The FBOS also contains a motion sensor. Motion sensor can detect abnormal shaking and a signal indicating car-crash will be sent back. In case of congestion or crash, car waits for instructions from control center. If the alarm is cleared, a 'resume' command will be sent. Otherwise, a 'manual' command informs the car to exit auto-parking mode and wait for manual control.

## IV. CONTROL CENTER DESIGN

### A. Overview

The other sub-system, control center, is designed to closely monitor the overall condition of the car park and the status of a parking process.

The control center consists of two parts, the ground station located in the car park and the Graphic User Interface in the control room. Real-time duplex communication channels are established. Information is exchanged between intelligent car and ground station via RF signal. Car status is reported to the ground station and forwarded to the GUI, while command from control room is sent to car by ground station. Linkage between ground station and GUI is established via cable connection using RS232 ports. An additional feature, communicates with car owners via SMS, is implemented in GUI using Bluetooth Technology. Fig.10 illustrates the communication channels between sub-systems.

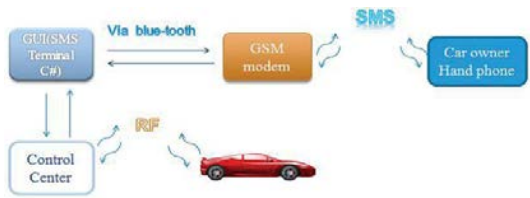

Fig. 10. Communication between Subsystems

### B. Functions and Realisation

*1) Real-time communication with car via RF:* Ground station and the car are equipped with RF signal transmitter and receiver modules, i.e. miniature UHF radio modules. These modules can establish a simple telemetry link up to 300 meters, and transmit data at a speed of up to 128KB/s. Wireless communication suffers a relatively high error rate due to various noises. Therefore, encryption is designed and implemented to ensure reliable data transmission. Each information packet consists of three bytes: address byte, data byte and check byte. Only host with the same address byte can access the data byte. Check byte is designed for further verification.

*2) RS232 serial port transmission:* Data transmission between ground station and PC in control room is achieved via cable connection. Data transmission and receipt in ground station is realized by universal synchronous-asynchronous receiver transmitter (USART) module. Asynchronous full duplex transmission is established between the two terminals, following RS232 serial communication standard. The PC serial port is connected with the microprocessor in ground station through a RS232 cable with the help of MAX233 chip. To minimize load and speed up transmission, status or command transmitted between the two sub-systems is encoded as a single alphabet (e.g. sending S indicating the car is in searching mode).

*3) Bluetooth communication:* Bluetooth technology is used to establish links between PC and GSM modem. GSM modem forwards the information to car owner as text messages. The information transmitted is a concatenated string containing the CarID, car location and other relevant information. This feature is integrated in GUI using C# programs. The particular library used to control GSM modem is Hayes command set, also known as AT commands. To prevent information leakage, encryption is implemented.

## V. IMPLEMENTATION AND EXPERIMENTAL RESULTS

To demonstrate that the designed intelligent autonomous parking system is feasible, a car prototype (Fig.11) and a model car park were constructed for test. The system is built on a scale of 1:14. The prototype is 220mm in length, 180mm in width, and 160mm in height. Dimensions of the car park (Fig.12) is 3m x 2.5m. Car park model consisted of two rows, one for parallel parking and the other for vertical parking. The ground station was located in the middle of the car park. The Fuzzy-Based Onboard System (FBOS) was implemented with the following components: PIC 16F877 16-bit microcontroller, ArduIMU, Sharp infra-red distance sensors (GP2D120XJ00F), DC motor with L293E motor driver, DC servo unit, R5-434-5 RF transmitter/receiver module pair, rotary encoder and motion sensor. The controller in ground station was implemented using PIC 16F877 16-bits microcontroller, R5-434-5 RF transmitter/receiver module pair, RS232 serial port with MAX233CPP RS232 driver, 16-bits LCD with backlight, membrane-type Keypad, buzzer and LED light.

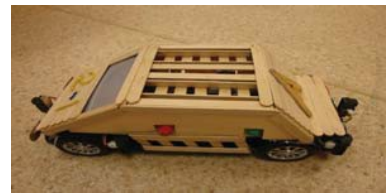

Fig. 11. Intelligent Car Model

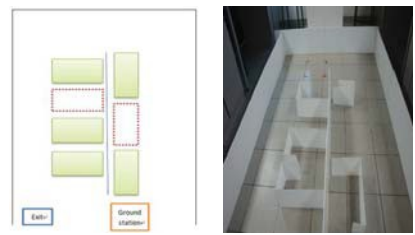

Fig. 12. Car park

The experiments tested the following functions: auto-driving, slots detection, parallel parking, vertical parking, and emergency handling.

Communication between different sub-systems are also tested. During the test, the ground station received the car status and all data was reflected on the LCD screen and GUI in PC terminal. Fig.13 illustrates the GUI during post-parking period. The upper left dialogue box is used to set up Bluetooth and serial port connection. Lower left part displays car location. The rest is an overview of the car park for surveillance.

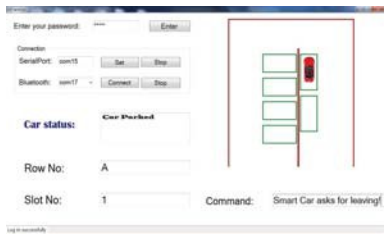

Fig. 13. Car parked status

The following images are extracted from the experiment video. Fig.14 demonstrated the whole procedure of parallel parking. Fig.15 demonstrated a successful vertical parking experiment. These experiments proved that the FBOS was robust to select the appropriate parking mode in different situations.

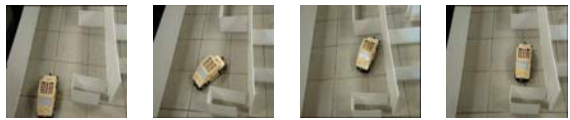

Fig. 14. Parallel parking

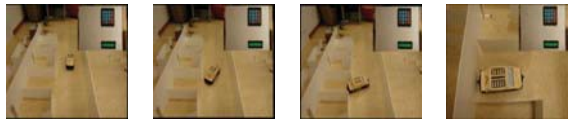

Fig. 15. Verticall parking

If an obstacle is detected in the parking process, the car stops and waits. Congestion is reported if obstacle is not removed in time. Location of congestion is sent back to control center through real-time RF communication and reflected on GUI. Fig.16 illustrates the car-park condition and GUI response in a congestion.

## VI. CONCLUSIONS AND FUTURE WORKS

In this paper, an intelligent autonomous parking system is proposed. The system consists of two sub-systems, Fuzzy-Based Onboard System (FBOS) and the control center. FBOS controls slot detection and autonomous parking, facilitated by other components like infra-red sensors and ArduIMU. The FBOS manages other supporting functions such as obstacle avoidance, communication with control center etc. The control center, consisting of ground station and GUI is designed for the purpose of real time monitoring. Communication channels through RF signal, serial port and Bluetooth are established to achieve the task.

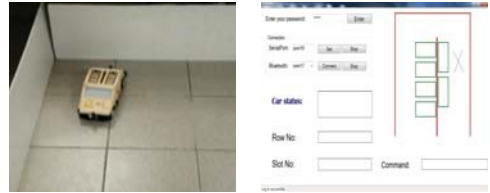

Fig. 16. Congestion mode

Future improvements should be focused on steering angle control. DC servo is used to control the steering angle of the front wheels. The servo draws power from voltage pulses whose duty cycle determines the steering angle. Ideally, the steering angle can be set as any value. However, fine tuning is prohibited by frictional forces and other noises. In the initial implementation, steering angle is limited to two values, i.e. small and large. Other possible methods and components for steering angle control should be explored in the future. An accurately controlled steering angle is essential to improve the performance.